\documentclass{aa} 

\usepackage{graphicx}
\usepackage{txfonts}
\usepackage{hyperref}
\usepackage{rotating}
\usepackage[author={RK}]{pdfcomment}

\graphicspath{{./images/}}

\newcommand{\glorentz}{{\gamma}}

\begin{document} 

\title{High-Resolution Simulations of LS~5039}

\author{
   R. Kissmann\inst{1}
   \and
   D. Huber\inst{1}
   \and
   P. Gschwandtner \inst{2}
}

\institute{
Institut f\"ur Astro- und Teilchenphysik,
Leopold-Franzens-Universit\"at Innsbruck,
6020 Innsbruck, Austria~\\
\email{ralf.kissmann@uibk.ac.at}
\and
Institut f\"ur Informatik,
Leopold-Franzens-Universit\"at Innsbruck,
6020 Innsbruck, Austria
}

\date{Received --; accepted --}

\abstract{
We present an analysis of our high-resolution relativistic-hydrodynamics model of the stellar- and pulsar-wind interaction in the LS-5039 system.
}{
With our high-resolution simulation covering three orbital periods, we analyse the impact of turbulence with a particular focus on short-term and orbit-to-orbit variations.
}{
Our model uses a relativistic hydrodynamics description of the wind interaction in the LS-5039 system assuming a pulsar-wind driven scenario.
The corresponding system of equations is solved using the finite-volume code \textsc{Cronos}.
We compute statistical quantities, also relevant for particle acceleration in this system, from results of multiple consecutive timesteps.
}{
In our simulation we find the previously observed shock structures related to the wind-collision region (WCR), including the pulsar-wind termination, being dynamically influenced by orbital motion.
In our high-resolution simulation we find high turbulence levels following from instabilities driven at the WCR.
These instabilities lead to strong fluctuations of several dynamical quantities especially around and after apastron.
These fluctuations are expected to impact the particle transport and also especially the related emission of non-thermal radiation.
As an important example, the region from which gamma-ray emission has been found to be boosted due to relativistic beaming in previous studies shows strong variations in size both on short and on orbital timescales.
}{
Using a large computational domain together with high spatial resolution allowed a detailed study of fluctuations in the stellar- and pulsar-wind interaction.
The results indicate a possible influence on the non-thermal emission from this system, which will be analysed with dedicated simulations in a forthcoming publication.
}


\maketitle

\section{Introduction} \label{sec_intro}
Gamma-ray binaries are binary systems which consist of a massive star and a compact object and emit the major part of their observable radiative output in the gamma-ray regime. A prominent approach to explain the observed non-thermal emission from some of these systems is the wind-driven scenario \citep[see][and references therein]{Dubus2013AnARv21_64}. In this case, the compact object is assumed to be a pulsar, where the highly relativistic pulsar wind forms a wind-collision region (WCR) when interacting with the massive stellar wind from the star. At the strong shocks surrounding the WCR, particles are thought to be injected and accelerated to very high energies \citep{MaraschiEtAl1981MNRAS194_1, Dubus2006AnA456_801}.

The most studied gamma-ray binary is the LS-5039 system, for which a plethora of observations is available at different energies.
Additionally, the properties of this system are comparatively well known: it contains an O-type star and a compact object in a mildly eccentric orbit around each other.
For this system, many studies assume the wind-driven scenario, where a small fraction of the rotational energy of the pulsar drives a pulsar wind interacting with the wind of the O-type star.
This scenario is supported by the orbital modulation observed in the non-thermal emission at different energies \citep[see, e.g.][]{Aharonian2005Sci309_746, Abdo2009ApJ706_56, Takahashi2009ApJ697_592}.
Additionally, there is tentative evidence of pulsations in X-rays from an analysis of Suzaku and Nustar data \citep{YonedaEtAl2020PhRvL125_1103}, which also hint at the presence of a pulsar as the compact object.
While \citet{Volkov2021ApJ915_61} discuss that the significance of the periodicity found by \citet{YonedaEtAl2020PhRvL125_1103} is rather low, they nonetheless argue from their observation of the lightcurve of LS~5039 that the colliding-wind scenario still is the most likely one.

In this scenario, the non-thermal emission from LS~5039 is strongly linked to the complex dynamics of the colliding pulsar and stellar winds within the WCR and beyond.
Correspondingly, this dynamical interaction has already been investigated in many numerical studies. 
Due to the high numerical cost of a fully relativistic description, early works focussed on individual aspects of such gamma-ray binary systems, where certain simplifications became feasible, like neglect of a relativistic description \citep{RomeroEtAl2007AnA474_15, TakataEtAl2012ApJ750_70}, neglect of orbital motion \citep{BogovalovEtAl2012MNRAS419_3426, LambertsEtAl2013AnA560_A79, DubusEtAl2015AnA581_27}, or simulations with reduced dimensionality \citep{Bosch-RamonEtAl2012AnA544A_59, BogovalovEtAl2012MNRAS419_3426, Paredes-Fortuny2015AnA574_77}.

First 3D, relativistic simulations of LS~5039 were discussed in \citet{Bosch-RamonEtAl2015AnA577_89}, where in particular the difference to 2D simulations and the evolution of instabilities were studied \citep[see also][for relativistic 3D simulation of other gamma-ray binaries]{Bosch-RamonEtAl2017MNRAS471_150, BarkovBosch-Ramon2016MNRAS456_64}.
By using a radially increasing cell size, the authors were able to simulate a larger computational domain, while still having high resolution near the apex of the WCR.
Like the previous two-dimensional study \citet{Bosch-RamonEtAl2012AnA544A_59}, \citet{Bosch-RamonEtAl2015AnA577_89} observed the spiral pattern caused by the orbital motion, which starts being disrupted at larger radii due to the turbulence driven in the WCR, as also observed in colliding-wind-binary simulations \citep[]{LambertsEtAl2012AnA546_60}.
However, due to the radially decreasing resolution, turbulence was attenuated in the outer parts of the simulations by \citet{Bosch-RamonEtAl2015AnA577_89}.
While \citet{HuberEtAl2021AnA649_71} used homogeneous resolution through their simulated domain, the authors found that the corresponding domain was too small to contain the unshocked pulsar wind at all orbital phases.

In this work, we extend the previous studies, by investigating a simulation of LS~5039 with homogeneous high spatial resolution throughout the entire numerical domain. 
In particular we increase the size of the numerical domain as compared to \citet{HuberEtAl2021AnA649_71} and further increase the spatial resolution to reduce the dissipation of turbulence in the simulation.
Additionally, we simulate the dynamics of the system for three full orbits, where the first orbit is only used to allow the system to settle into a quasi steady state. With this, we are able to observe the impact of turbulence also in the outer parts of the numerical domain and not only quantify the short-term variation but also the orbit-to-orbit variability of LS~5039.

The paper is structured as follows: in Sec. \ref{SecSetup} we introduce the specific setup of our simulation together with the mathematical description. The results are discussed in Sec. \ref{SecResults}, where we start by investigating the flow structure and the dynamics and end with an analysis of the short-term and orbit-to-orbit variations. Finally, we summarise our findings in Sec. \ref{SecSummary}.

\section{Physical and Numerical Setup}
\label{SecSetup}
Since we simulate the interaction of a highly-relativistic pulsar wind with the massive wind of an early-type star, we use relativistic hydrodynamics (RHD) to describe the wind dynamics \citep[for a more extensive discussion of the model see][]{HuberEtAl2021AnA646_91}. 
In particular, we solve the following set of partial differential equations:
\begin{align}
	\label{EqMassConservation}
	\frac{\partial D}{\partial t}
	+
	\nabla \cdot
	\left(
	\frac{1}{\glorentz} D \vec{u}
	\right)
	&= 0
	\\
	\label{EqEnergyConservation}
	\frac{\partial \tau}{\partial t}
	+
	\nabla \cdot
	\left(
	\frac{1}{\glorentz}
	\left( \tau + p \right)
	\vec{u}
	\right)
	&=
	\vec{S}_\tau
	\\
	\label{EqMomentumConservation}
	\frac{\partial \vec{m}}{\partial t}
	+
	\nabla \cdot
	\left(
	\frac{1}{\glorentz} \vec{m} \vec{u}
	\right)
	+
	\nabla p
	&= 
	\vec{S}_m
\end{align}
where we have the vector $\vec{U}$ of conserved quantities
\begin{equation}
	\vec{U} = \begin{pmatrix}
    D \\
    E \\
    \vec{m}
  \end{pmatrix}
  =
  \begin{pmatrix}
    \glorentz \rho \\
    \glorentz \rho \left( \glorentz h - 1 \right)\\
    D h \vec{u}
  \end{pmatrix}.
\end{equation}
Here, $\rho$ is the mass density, $\vec{u}$ is the spatial vector of the relativistic four velocity with $u^i = \gamma v^i$, $h$ is the specific enthalpy, and $p$ is the thermal pressure. 
Additionally, $\tau$ is related to $E$ via $E = \tau + D$.
With our normalisation $c=1$, $\vec{v}$ is given in units of the speed of light and we find for the relativistic Lorentz factor:
\begin{equation}
\glorentz = \frac{1}{\sqrt{1-\vec{v}^2}} = \sqrt{1 + \vec{u}^2}
\end{equation}
i.e., the Lorentz factor can be directly computed from the spatial components of the relativistic four velocity.

The system of equations given in Eqs. \ref{EqMassConservation} - \ref{EqEnergyConservation} is closed by an ideal equation of state:
\begin{equation}
  h(\rho, p) = 1 + \frac{\Gamma}{\Gamma - 1} \frac{p}{\rho},
\end{equation}
where we use a constant adiabatic exponents $\Gamma = 4/3$ \citep[see][for a related discussion]{HuberEtAl2021AnA646_91}.
Here, we did not solve an equation for $E$ as is often done in other simulations \citep[see, e.g.,][]{MignoneEtAl2007ApJS170_228}, since $D$ and $E$ can become very similar in the low-pressure, non-relativistic regime.
With both $D$ and $E$ being conserved quantities, $\tau$ is also a conserved, which is more suitable in our case especially for the treatment of the stellar wind.
Additionally, we solved a supplementary conservation equation for the specific entropy $s = p / \rho^\Gamma$, which is only conserved in smooth flows, and can there be used to overcome possible problems with negative pressure \citep[see][]{HuberEtAl2021AnA646_91}.

These equations were solved using the \textsc{Cronos} code \citep[][]{KissmannEtAl2018ApJS236_53}, which was recently extended to allow simulations of relativistic hydrodynamics \citep[see][]{HuberKissmann2021AnA653_164}. In this work, we applied piecewise linear spatial reconstruction using the minmod limiter \citep[][]{VanLeer1979JCoPh32_101, Harten1983SIAMReview25_35} together with an \textsc{Hllc} RHD Riemann solver \citep[][]{MignoneBodo2005MNRAS364_126}. The simulation analysed in this work was run on the  Joliot-Curie system at GENCI@CEA, France.

\subsection{Simulation Setup}
In our model of LS~5039, we simulated the wind dynamics of the system, consisting of a pulsar emitting a relativistic pulsar wind in orbit with a massive O-type star, for three full orbits.
Here, we use the same physical parameters for the system as given in  \citet{HuberEtAl2021AnA649_71}: we adopt the orbital parameters from \citet{CasaresEtAl2005MNRAS364_899} using a period of $P_\text{orbit} = 3.9$~d, and eccentricity of $e=0.35$, and masses of $M_{\text{star}}=23 M_{\sun}$ and $M_{\text{pulsar}} = 1.4 M_{\sun}$ \citep[see also][]{Dubus2013AnARv21_64}.

As discussed in the introduction, we increased both the extent of the numerical domain and also the spatial resolution in comparison to our previous simulations of LS~5039.
With the center of mass at the coordinate origin we use a numerical domain with dimensions $[-2,2] \times [-0.5,2.5]\times[-1,1]$ AU$^3$ which is homogeneously filled with 2048$\times$1536$\times$1024 cubic cells.
In total this leads to a resolution of \mbox{$\sim$0.002 AU} or \mbox{$\sim 0.42$ $\text{R}_\odot$} in each spatial dimension, i.e. about the same resolution as used in the inner region by \citet{Bosch-RamonEtAl2015AnA577_89} or a doubling of spatial resolution together with a significant increase of the simulation volume in comparison to the previous study by \citet{HuberEtAl2021AnA649_71}.
In the present study, we use homogeneous resolution throughout the numerical domain.
Adaptive mesh refinement was not considered, since small-scale fluctuations filled most of the computational volume \citep[see also][]{Bosch-RamonEtAl2015AnA577_89}.

In our simulation, the $z$-axis is perpendicular to the orbital plane. 
As before, the simulation was performed in a reference frame co-rotating with the average angular velocity of the system, which allowed to use a smaller non-symmetric domain than in a non-rotating setup, which contains the full unshocked pulsar wind at all times of the simulation.
By using the so-called 3+1 Valencia formulation of general relativistic hydrodynamics \citep[][]{Banyuls1997ApJ476_221}, this requires using the source term:
\begin{equation}
  {\bf S_m} = \left( \Omega \, m_{y}, \, -\Omega \, m_{x}, \, 0 \right)^\top
\end{equation}
where $\Omega$ is the average angular velocity of the binary \citep[see][for further details]{HuberEtAl2021AnA646_91}.
Since we are dealing with an eccentric orbit in the case of LS~5039, the two stars still show residual motion in the system rotating with the average angular velocity.
This is apparent by the axis connecting the binaries not being aligned with the coordinate axes for most of the phases depicted, e.g., in Fig. \ref{FigDensOrbit2}.
Periastron is marked by phase $\Phi=0$, where the axis connecting the binaries is aligned with the $y$-axis of our simulation volume.

In the present case, the simulation was performed for 3 full orbits of the system, where the analysis focusses on the second and the third orbit (the first orbit was used to allow the system to change from the initial homogeneous low-density medium to the fully turbulent state).
Here, we allowed more time for this initialisation phase than in our previous study because of the larger spatial domain.
In the present simulation the undisturbed wind of the massive star reaches the farthest corner of the numerical domain at $t_\text{crossing} \simeq 0.75 P_{\text{orbit}} \simeq 2.9$~d. 
This is actually a more conservative estimate than in \citet{Bosch-RamonEtAl2015AnA577_89}, who initialised most of the domain with the expanding stellar wind, allowing $\simeq 0.5$~d for initialisation.

For the stellar wind, we used a stellar mass-loss rate of $\dot{M}_s = 2 \cdot 10^{-8} M_{\sun}$ yr$^{-1}$ together with a terminal velocity of $v_s=2000$ km/s \citep[][]{DubusEtAl2015AnA581_27}.
The stellar wind was injected as an isotropic outflow in a sphere around the position of the star with radius $r_{\text{inj}} = 0.012$~AU, i.e., with a radius of 6 computational cells, where the star moves in the co-rotating frame due to its eccentric orbit.
The pulsar wind was similarly injected with a speed of $v_p = 0.99 c$, with $c$ the speed of light.
For the pulsar's mass-loss rate we used $\dot{M}_p = \eta \frac{\dot{M}_s v_s}{u_p}$ with $\eta=0.1$ as also used in \citet{Bosch-RamonEtAl2015AnA577_89}.
With this, we assume a pulsar spin-down luminosity of $L_{SD} = 7.55\cdot 10^{28}$~W.
The thermal pressure at the outer radius of the injection sphere was set to be a fraction of $10^{-9}$ and $10^{-7}$ of the local rest-energy density for the star and the pulsar, respectively.

For analysis we stored 20 output files evenly distributed over orbital phase for each orbit.
For a little more than half of these outputs after the first orbit, we produced a series of six successive output datasets with only about $8.3$ min between the individual datasets.
These can be used to investigate short-term fluctuations and compute statistics of relevant quantities.

\section{Results}
\label{SecResults}
\begin{figure*}
\centering
\setlength{\unitlength}{0.0166666cm}
\begin{picture}(1020,598)(-20,-30)
\put(-20,120){\small \rotatebox{90}{$y$ [AU]}}
\put(-20,350){\small \rotatebox{90}{$y$ [AU]}}
\put(180,-20){\small $x$ [AU]}
\put(485,-20){\small $x$ [AU]}
\put(790,-20){\small $x$ [AU]}
\put(480,545){\small $\rho$ [kg m$^{-3}$]}

\includegraphics[width=1000\unitlength]{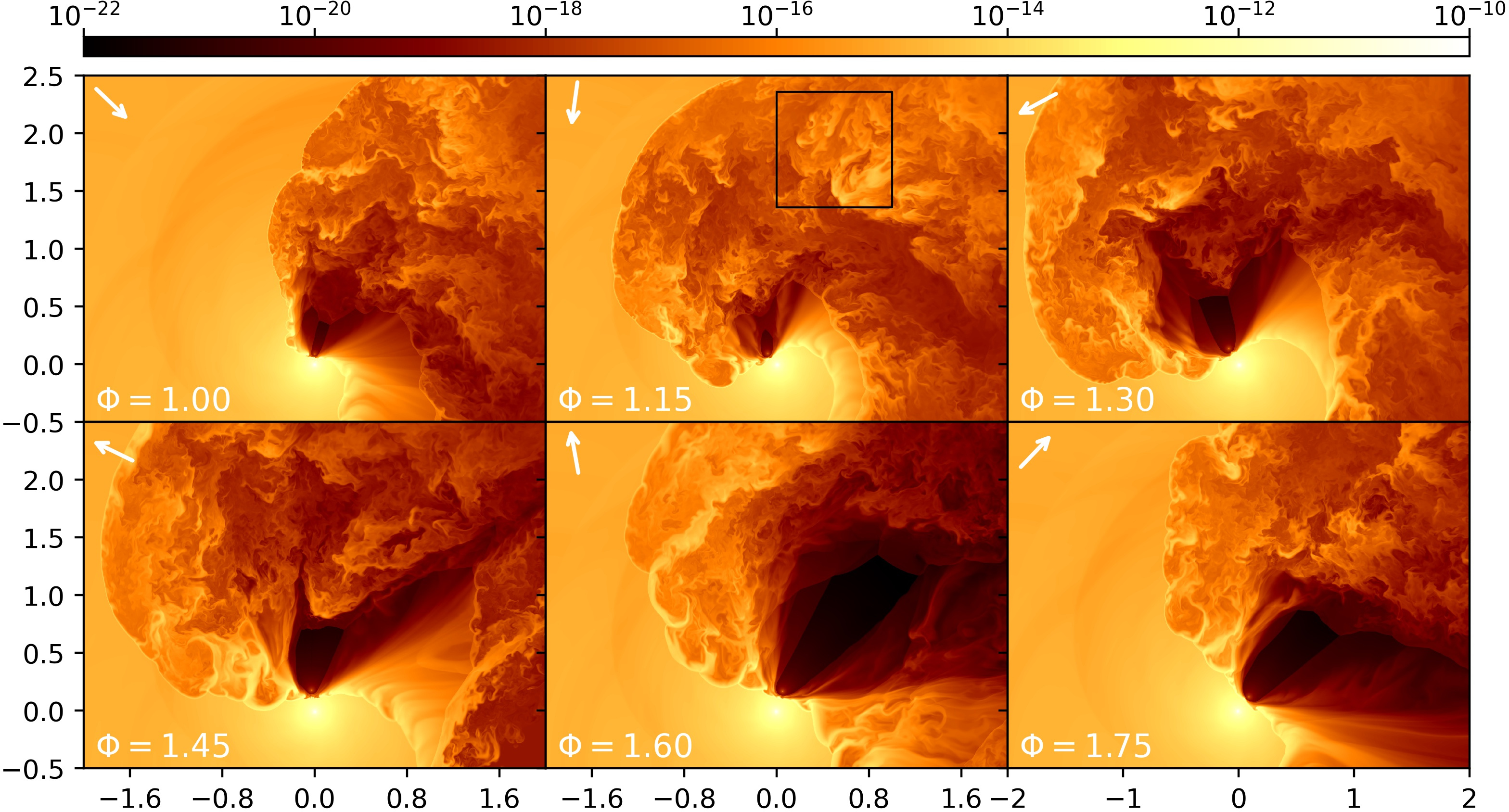}
\end{picture}
\caption{Mass density in the orbital plane during the second full orbit. Corresponding phases are given in the individual sublots together with an arrow towards the observer location. The center of gravity is at $x=0, y=0$.
For the central plot at the top, the black box indicates the region, for which we performed the turbulence analysis.}
\label{FigDensOrbit2}
\end{figure*}
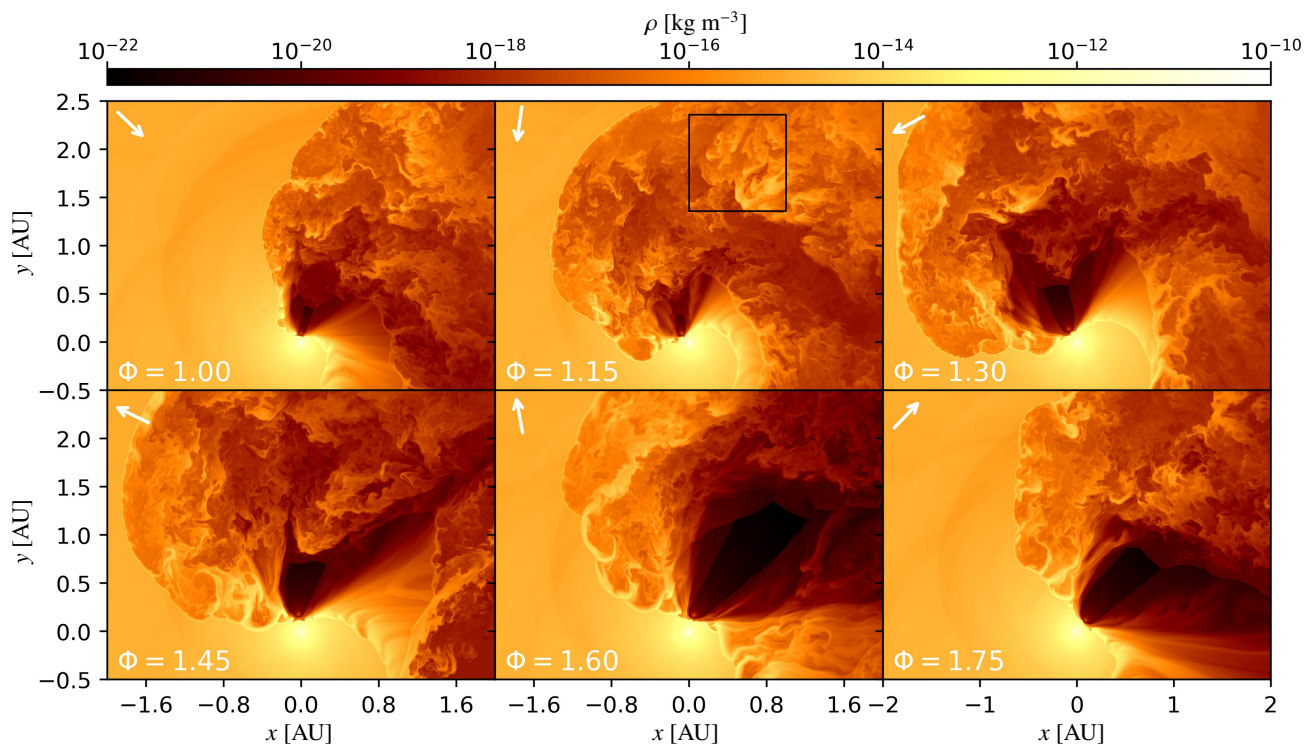

In Fig. \ref{FigDensOrbit2} we show the density in the orbital plane for six different phases of the second simulated orbit.
In these figures, the dark, mostly featureless region is the radially expanding supersonic pulsar wind, where the bright dot marks the local density peak at the location of the pulsar.
The homogeneous region to the left / lower left contains the unshocked, radially expanding stellar wind, where the star is visible as the large bright circle near $(x,y) = (0,0)$. Between the star and the pulsar, one can see the wind-collision region (WCR), where both winds interact.
On both sides, this WCR is bounded by a shock, where the wind-component parallel to the shock normal becomes subsonic.
For the stellar wind, this termination shock shows a spiral pattern owing to the orbital motion of the stars.
This becomes even more apparent at larger scales as investigated in \citet{Bosch-RamonEtAl2015AnA577_89}.
The termination shock of the pulsar wind can be split into a u-shaped bow shock and a so-called Coriolis shock formed due to the orbital motion of the system \citep[][]{Bosch-RamonBarkov2011AnA535_20, Bosch-RamonEtAl2012AnA544A_59}, where the latter is also a possible site for particle acceleration \citep[][]{Zabalza2013AnA551_17, HuberEtAl2021AnA649_71}.

In the downstream regions of the WCR, Fig. \ref{FigDensOrbit2} shows space-filling turbulence, that is driven by different instabilities within the WCR.
On the one hand, the shear-flow between the shocked stellar and pulsar winds at the contact discontinuity within the WCR triggers Kelvin-Helmholtz (KH) instabilities \citep[][]{Bosch-RamonEtAl2012AnA544A_59, LambertsEtAl2013AnA560_A79, Bosch-RamonEtAl2015AnA577_89}.
Additionally, the Richtmyer-Meshkov \citep{Richtmyer1960CommPAM13_297, Meshkov1972FlDy4_101} (RM) and the Rayleigh-Taylor (RT) \citet{Rayleigh1882ProcLonMathSoc_s1_170, Taylor1950RSPSA201_192} instabilities are active in this highly dynamical environment, where \citet{Bosch-RamonEtAl2015AnA577_89} discuss that they act together as an important driver of the turbulence in these systems.
As is also visible in our Fig. \ref{FigDensOrbit2} turbulence is much stronger at the leading edge of the WCR (to the left of the cavity containing the unshocked pulsar wind).

  This is a strong indication that the RT instability is acting as one of the important drivers of turbulence in these systems, because only at the leading edge do acceleration (by the Coriolis force) and density gradient point in the opposite direction \citep[see also][]{Bosch-RamonEtAl2015AnA577_89}.
  While \citet{GuzdarEtAl1982GeoRL9_547} point out that a velocity shear will modify the RT instability, they only find a significant reduction of the growth rate for wavelengths shorter than the scale of the density gradient, which in our case is the grid scale.
  At large scales, where the relevant driving occurs in our simulations, the RT instability is expected to be unaffected by the velocity shear at the contact discontinuity.

The presence of the RT and the RM instabilities is particularly important,
since our simulation cannot capture ultra-relativistic Lorentz factors as conventionally expected for pulsar winds \citep[][]{Aharonian2012Natur482_507, Khangulyan2012ApJ752_17}.
Since larger Lorentz factors lead to a correspondingly smaller mass-loss rate $\dot{M}_p$ for the pulsar and thus a larger density contrast, we expect longer growth timescales for the KH instability \citep[][]{FerrariEtAl1980MNRAS193_469}, which is also observed in the classical limit \citep[see, e.g.,][]{Chandrasekhar1961Book}.
In contrast, the growth rate of the RT instability depends on the difference of the densities of the stellar and the pulsar wind 
  \citep[see, e.g., Eq. (7) in][]{AllenHughes1984MNRAS208_609}, where the density of the pulsar wind is already negligible in our setup and even more so for larger Lorentz factors.
Thus, we expect little change for higher Lorentz factors for the RT instability.

\label{SecCoriolisShock}
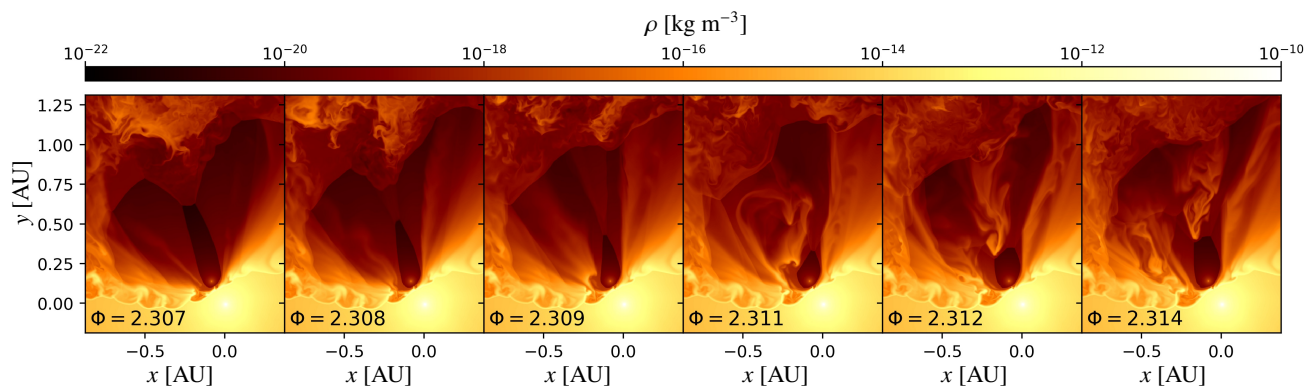
\begin{figure*}
	\centering
	\setlength{\unitlength}{0.0166666cm}
	\begin{picture}(1020,309)(-20,-30)
		\put(-20,105){\small \rotatebox{90}{$y$ [AU]}}
		\put(480,260){\small $\rho$ [kg m$^{-3}$]}
		\put(86,-20){\small $x$ [AU]}
		\put(243,-20){\small $x$ [AU]}
		\put(400,-20){\small $x$ [AU]}
		\put(556,-20){\small $x$ [AU]}
		\put(713,-20){\small $x$ [AU]}
		\put(870,-20){\small $x$ [AU]}
		\includegraphics[width=1000\unitlength]{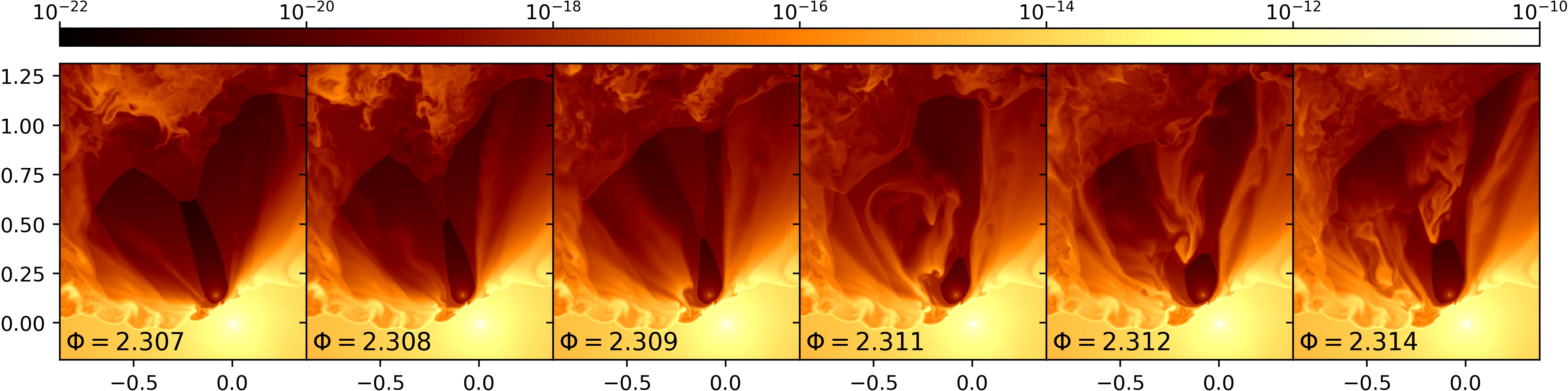}
	\end{picture}
	\caption{Different snapshots of mass density in the orbital plane during a short period in the second full orbit as indicated in the plots. As given by the indicated orbital phases, these snapshots cover a period of slightly more than 40 min.}
	\label{FigShortTermVariation}
\end{figure*}

Here, we feature a larger Lorentz factor than \citet{Bosch-RamonEtAl2015AnA577_89}, where the turbulence still shows the same general structure in our simulation: we also find ubiquitous turbulence and corresponding mixing downstream of the WCR.
Also, in our simulation, we find that the instabilities within the WCR drive the mass loading, i.e. dense matter from the stellar wind is mixed with the dilute, high-velocity pulsar-wind material \citep[see also][]{Bosch-RamonEtAl2012AnA544A_59}.
This is visible by the dense clumps of shocked stellar-wind material embedded in this region, where in our case the mixing already seems stronger near the apex of the WCR than observed by \citet{Bosch-RamonEtAl2015AnA577_89}.
Additionally, the fluctuations of the WCR are much stronger than in \citet{Bosch-RamonEtAl2015AnA577_89}, where in particular the size of the region containing the unshocked pulsar wind varies more strongly.
This difference can be attributed at least to two effects: first, we use a slightly larger eccentricity in our simulation, and secondly, due to the homogeneous resolution in our simulation, turbulence levels in the outer parts of the numerical domain are not damped and can therefore lead to a disruption of the structure of the supersonic pulsar wind volume.

\subsection{Coriolis-shock region}
\label{SecCoriolisRegion}
The position and structure of the Coriolis shock in our simulation is subject to dynamical changes, as visualised in Fig. \ref{FigShortTermVariation}. There, we show a zoom into the region containing the unshocked pulsar wind over a period of slightly more the 40 min. Apparently, the distance of the Coriolis shock from the apex of the WCR changes significantly, where it is absent for parts of that period, i.e., the flanks of the WCR converge into a point instead of connecting to both ends of the Coriolis shock. This will also have implications for particle acceleration at this shock, where a change in distance from the star will also change the energy loss rates of particles at the shock.

\begin{figure*}
	\centering
	\setlength{\unitlength}{0.0163cm}
	\begin{picture}(1040,275)(-20,-30)
		\put(-20,105){\small \rotatebox{90}{$y$ [AU]}}
		\put(1010,120){\small \rotatebox{90}{$u$}}
		\put(160,-20){\small $x$ [AU]}
		\put(455,-20){\small $x$ [AU]}
		\put(750,-20){\small $x$ [AU]}
		\includegraphics[width=1000\unitlength]{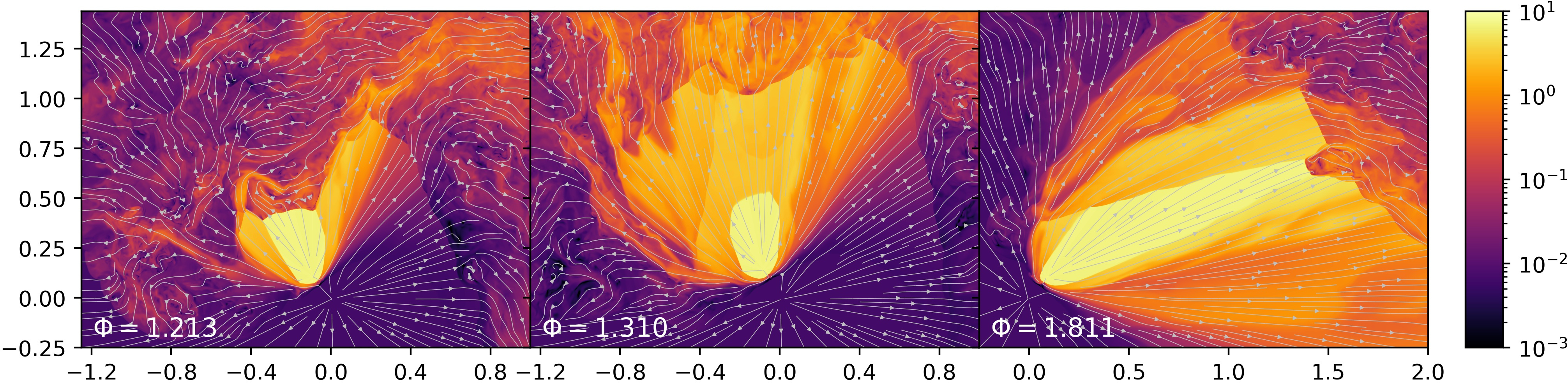}
	\end{picture}
	\caption{Absolute value of the spatial component of the fluid's four velocity in the orbital plan at selected phases as indicated in the plots. Superimposed over the images, we indicate the direction of the flow via streamlines. Here, we show a zoom into the computation domain focussing on the region around the Coriolis shock. Be aware of the different $x$-axis in the right-handed plot.}
	\label{FigVelocitySnapshots}
\end{figure*}
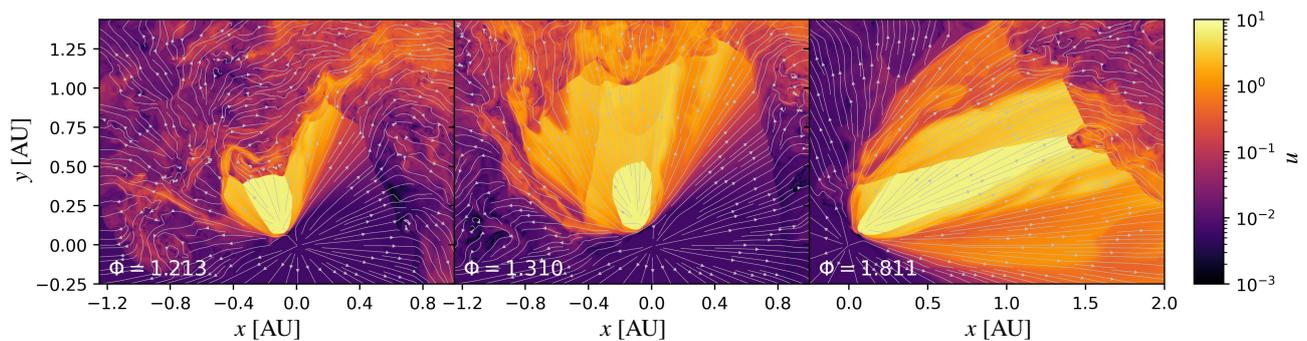
Further peculiarities of the shocks in the wind-collision region are visible in Fig. \ref{FigVelocitySnapshots}, where we show the absolute value of the spatial component of the relativistic four velocity of the fluid $u = v \gamma$.
Apparently, sometimes the Coriolis shock marks the transition to a region with the highly turbulent medium containing a mixture of stellar and pulsar wind, while at other phases it is embedded within the relativistic fluid from the wings of the WCR propagating between the unshocked pulsar wind and the contact discontinuity, or in a mixed state (right plot in Fig. \ref{FigVelocitySnapshots}).

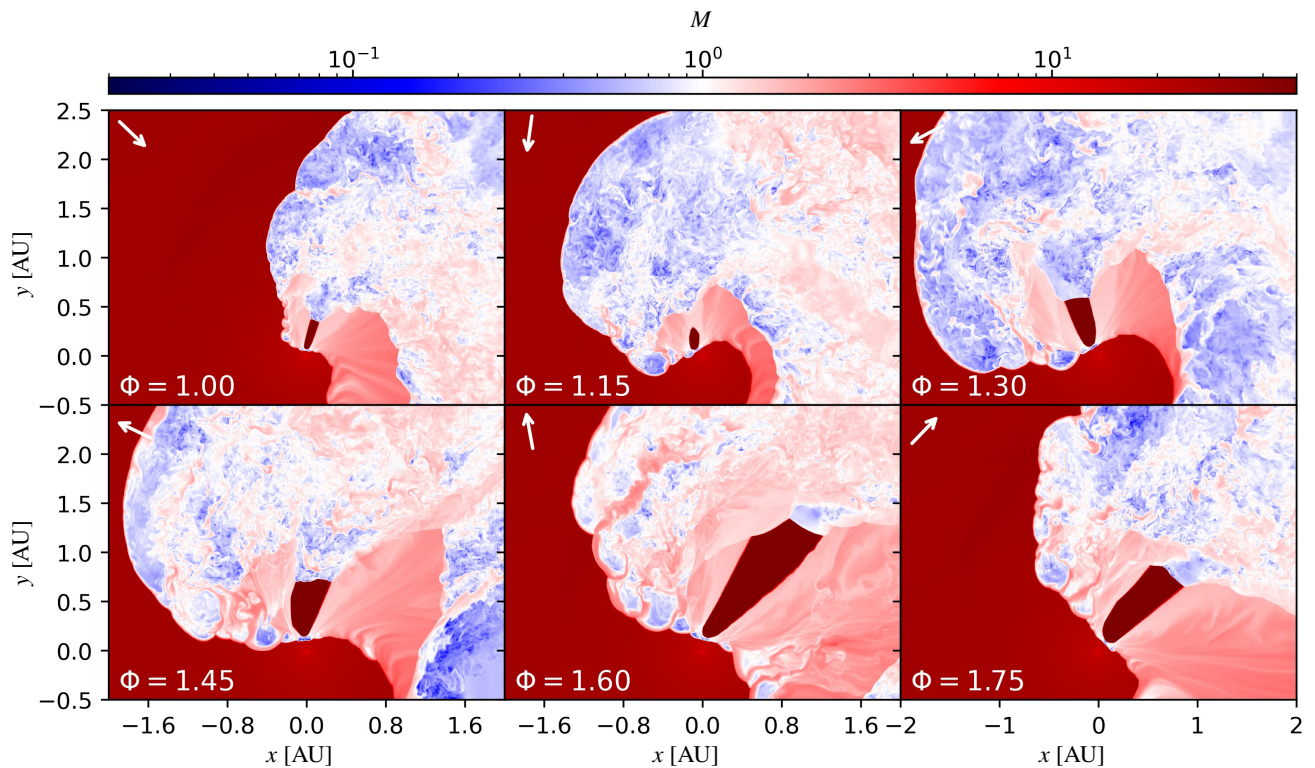
\begin{figure*}
\centering
\setlength{\unitlength}{0.0166666cm}
\begin{picture}(1020,609)(-20,-30)
\put(-20,120){\small \rotatebox{90}{$y$ [AU]}}
\put(-20,350){\small \rotatebox{90}{$y$ [AU]}}
\put(180,-20){\small $x$ [AU]}
\put(485,-20){\small $x$ [AU]}
\put(790,-20){\small $x$ [AU]}
\put(515,565){\small $M$}

\includegraphics[width=1000\unitlength]{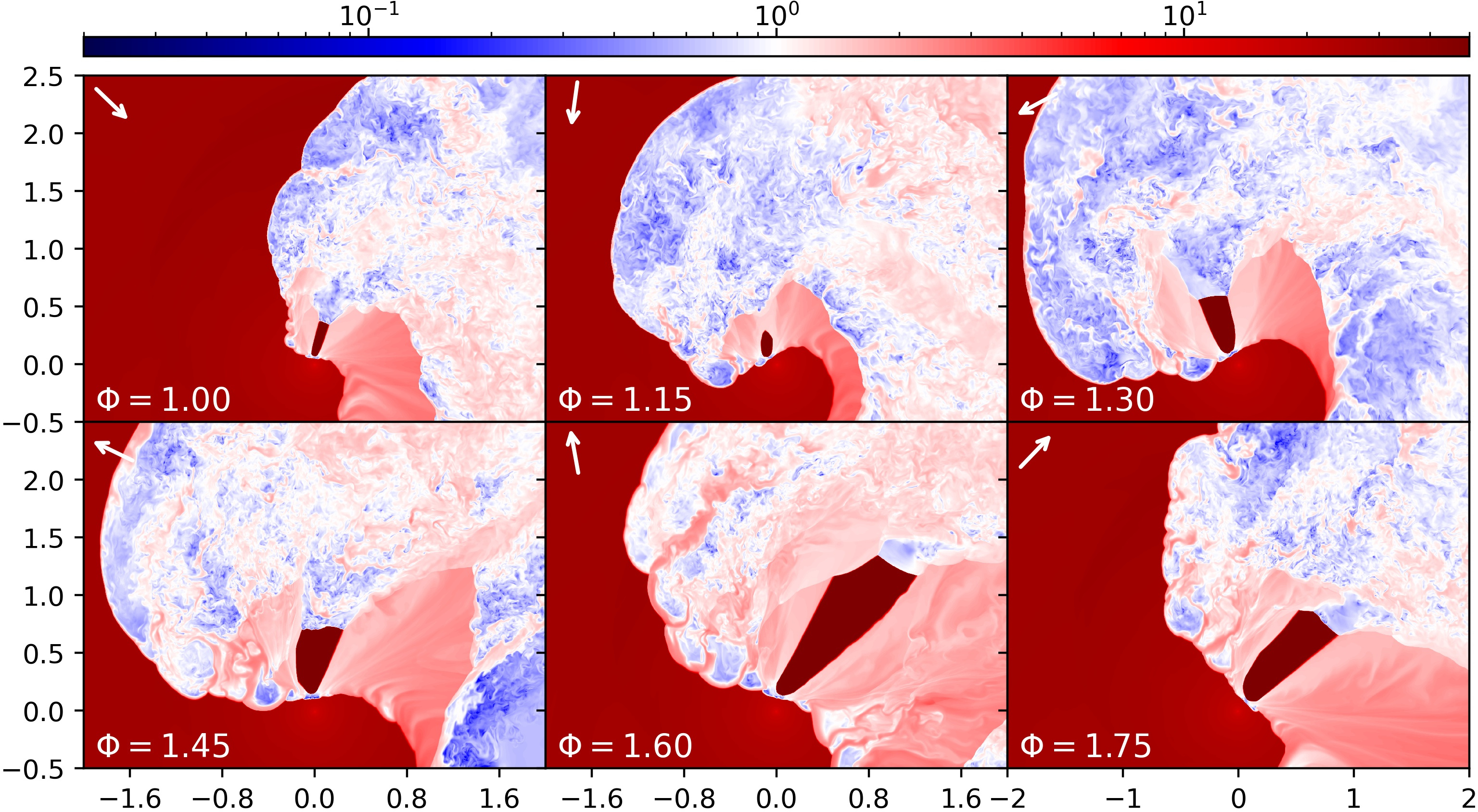}
\end{picture}
\caption{Mach number in the orbital plane during the second full orbit, for orbital phases as indicated in the subplots.
The arrows indicate the direction towards the observer.}
\label{FigMachOrbit2}
\end{figure*}

This rather laminar fluid in the wings of the WCR, contains streamlines connecting to the apex of the WCR, where the fluid is re-accelerated by the large pressure gradient \citep{BogovalovEtAl2008MNRAS387_63}, and streamlines connecting to the flank of the pulsar-wind termination shock, where the velocity component perpendicular to the shock normal is still highly relativistic -- in total we find Lorentz factors up to $\sim$3 in this region.
In all cases, this flow also terminates at an extended shock, when entering the region of highly turbulent, mixed matter downstream of the WCR.
The material in this region of mostly laminar flow correspondingly still moves supersonically as can be seen in Fig. \ref{FigMachOrbit2}, where we plot the Mach number in the orbital plane during the second full orbit.
The highly turbulent matter beyond the extended shock finally features large fractions of subsonic gas.
Also beyond the curved termination shock of the stellar wind, we see large regions of turbulent low-Mach-number flow.
Only the trailing-edge region beyond the stellar-wind termination shock features supersonic flow that later terminates in an extended shock.
The turbulence beyond these extended shocks is also obvious via the strong fluctuation in the Mach number.

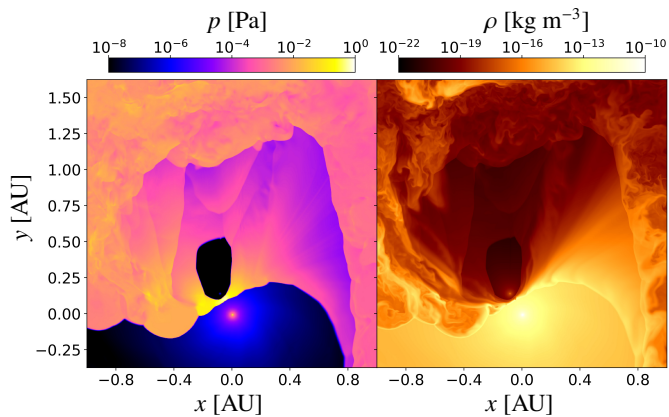
\begin{figure}
\resizebox{\hsize}{!}{
	\setlength{\unitlength}{0.0081cm}
	\begin{picture}(1050,653)(-50,-50)
		\put(-40,230){\small \rotatebox{90}{$y$ [AU]}}
		\put(250,-40){\small $x$ [AU]}
		\put(710,-40){\small $x$ [AU]}
		\put(710,570){\small $\rho$ [kg m$^{-3}$]}
		\put(270,570){\small $p$ [Pa]}
		\includegraphics[width=1000\unitlength]{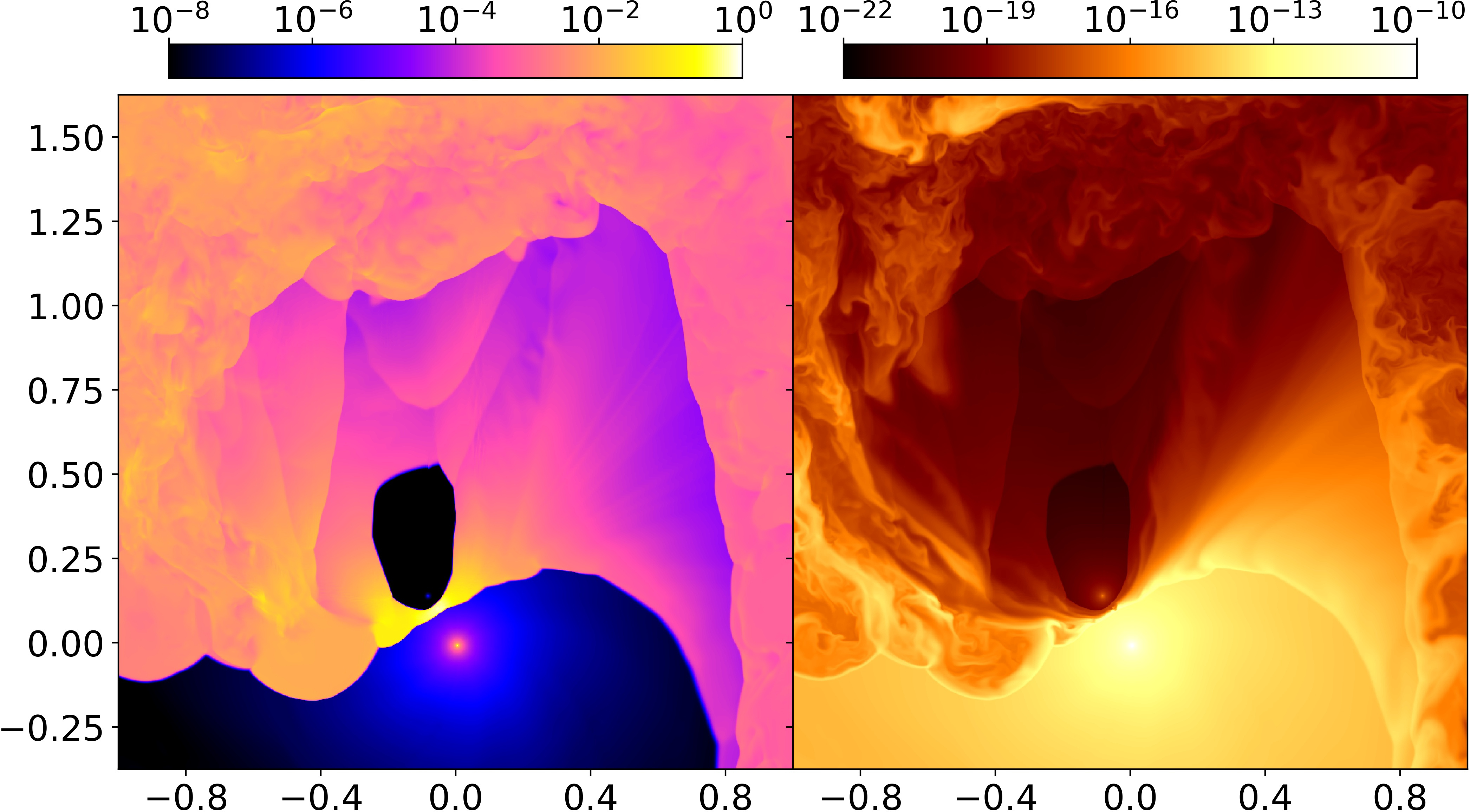}
	\end{picture}
	}
	\caption{\label{FigCoriolisPressure}Pressure (left) and mass density (right) in the region surrounding the unshocked pulsar wind for phase $\Phi=1.310$.}
\end{figure}
At some times, the region of laminar shocked material is also permeated by large-scale discontinuities, as visible at orbital phases $\Phi=1.310$ and $\Phi=1.811$ in Fig. \ref{FigVelocitySnapshots}.
As an example, we show mass density and thermal pressure for $\Phi=1.310$ in Fig. \ref{FigCoriolisPressure}, which feature the same large-scale discontinuity as the flow velocity.
Apparently, these discontinuities are shock waves connected to eddies in the contact discontinuity separating the shocked stellar wind from the shocked pulsar wind.
  By visual inspection it seems that the fluctuations increase beyond the point, where the large-scale shocks traverse the contact discontinuity.
  Since we do not have outputs at sufficiently short time intervals available to analyse the short-timescale evolution of these fluctuations, this can only be viewed as a hint that the RM instability might be responsible at least for parts of the fluctuations visible in our simulations \citep[see also][]{Bosch-RamonEtAl2015AnA577_89}.
  Apart from that, it
will be interesting to see how relevant these shock structures are for acceleration and the subsequent non-thermal emission, which has not been studied so far.

\subsection{Downstream Flow and Turbulence}
\label{SecFlowAndTurbulence}
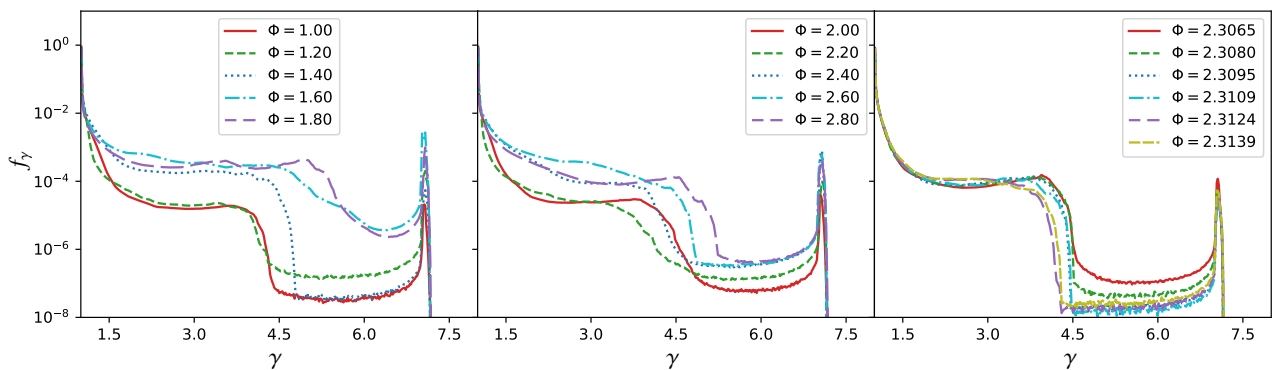
\begin{figure*}
	\centering
	\setlength{\unitlength}{0.0163cm}
	\begin{picture}(1040,298)(-20,-30)
		\put(-20,138){\small \rotatebox{90}{$f_{\gamma}$}}
		\put(190,-20){\small $\gamma$}
		\put(511,-20){\small $\gamma$}
		\put(832,-20){\small $\gamma$}
		\includegraphics[width=1000\unitlength]{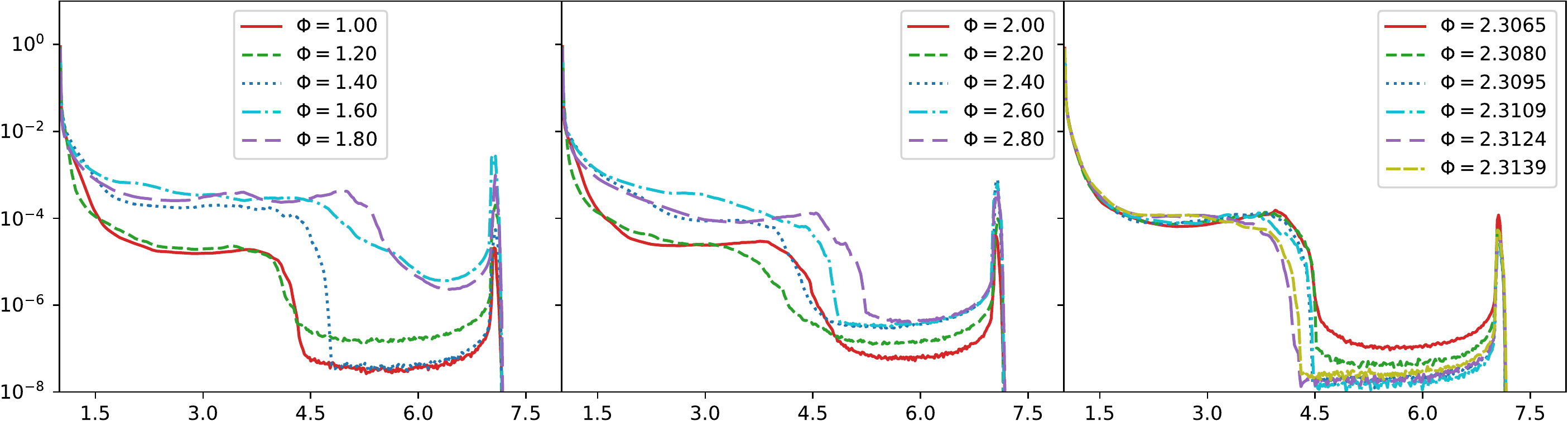}
	\end{picture}
	\caption{\label{FigDistGamma}Distribution function of the relativistic gamma factor $f_{\gamma}$ within the orbital domain at different phases as indicated in the plots. Here, we show results for the second orbit (left), for the third orbit (middle), and also for the same short time interval as visualised in Fig. \ref{FigShortTermVariation} (right).}
\end{figure*}

Beyond the extended shock structures we observe a highly turbulent downstream region embedded on the left in the unshocked stellar wind.
Here, the mixing of stellar- and pulsar-wind material is obvious both in Figs. \ref{FigDensOrbit2} and \ref{FigVelocitySnapshots}, where we observe high density clumps from the stellar wind or regions with a high Lorentz factor from the pulsar wind. 
In our case, this mixing shows up already close to the apex of the WCR, showing the efficiency of the instabilities acting to produce the turbulence.
Corresponding statistics is visible in Fig. \ref{FigDistGamma}, where we show the distribution function of the Lorentz-factor within the entire numerical domain for different orbital phases.

Given the high resolution together with the homogeneous grid in our simulation, we also investigated the longitudinal structure functions as a measure for the turbulence in our computational domain.
Since only the downstream medium shows strong fluctuations, we restricted the corresponding analysis to this region. 
In particular, we extracted the spatial part of the relativistic four-velocity field from a region with an extent of 1~AU in all spatial dimensions, giving a 512$\times$512$\times$512 data cube. 
This subdomain is centred around the orbital plane and located at $x=0\dots1$~AU, $y=1.36\dots2.36$~AU, as indicated via the black box in Fig. \ref{FigDensOrbit2}.
The longitudinal structure functions were computed according to \citep[see, e.g.][]{ZrakeMacFadyen2013ApJ763_12}
\begin{equation}
S_p^{\parallel} (l) = \left< \left| \frac{1}{l} \delta u^{\mu} \delta x_{\mu} \right|^p \right>
\end{equation}
where $\delta u^{\mu} = u_2^{\mu} - u_1^{\mu}$ and $l=\left(\delta x^{\mu} \delta x_{\mu}\right)^{1/2}$ with $\delta x^{\mu} = x^{\mu}_2 - x^{\mu}_2$ the separation four vector between pairs of points chosen randomly.
Here, we selected points that were simultaneous in the co-rotating frame.

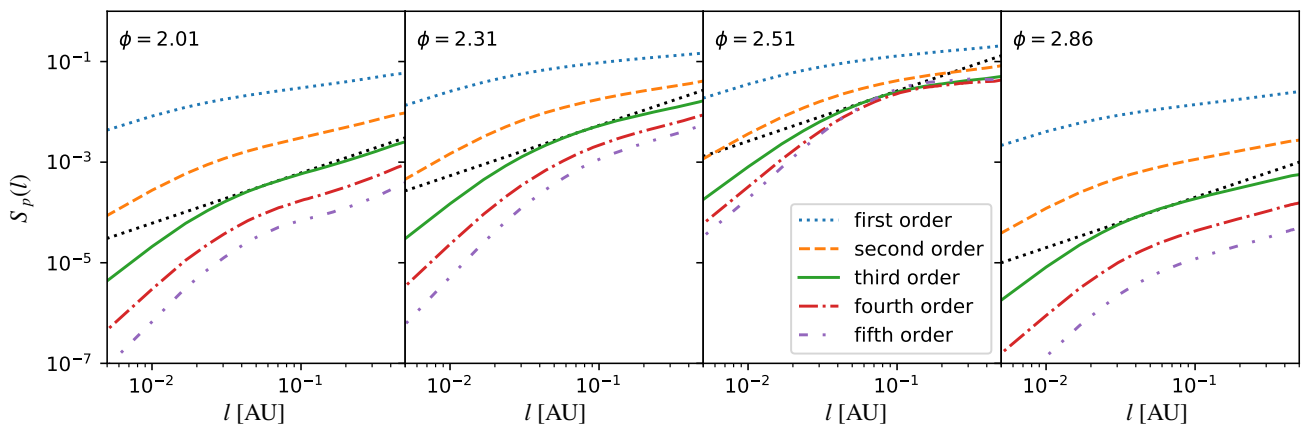
\begin{figure*}
	\centering
	\setlength{\unitlength}{0.0165cm}
	\begin{picture}(1030,330)(-30,-20)
		\put(-30,138){\small \rotatebox{90}{$S_p(l)$}}
		\put(140,-20){\small $l$ [AU]}
		\put(380,-20){\small $l$ [AU]}
		\put(620,-20){\small $l$ [AU]}
		\put(860,-20){\small $l$ [AU]}
		\includegraphics[width=1000\unitlength]{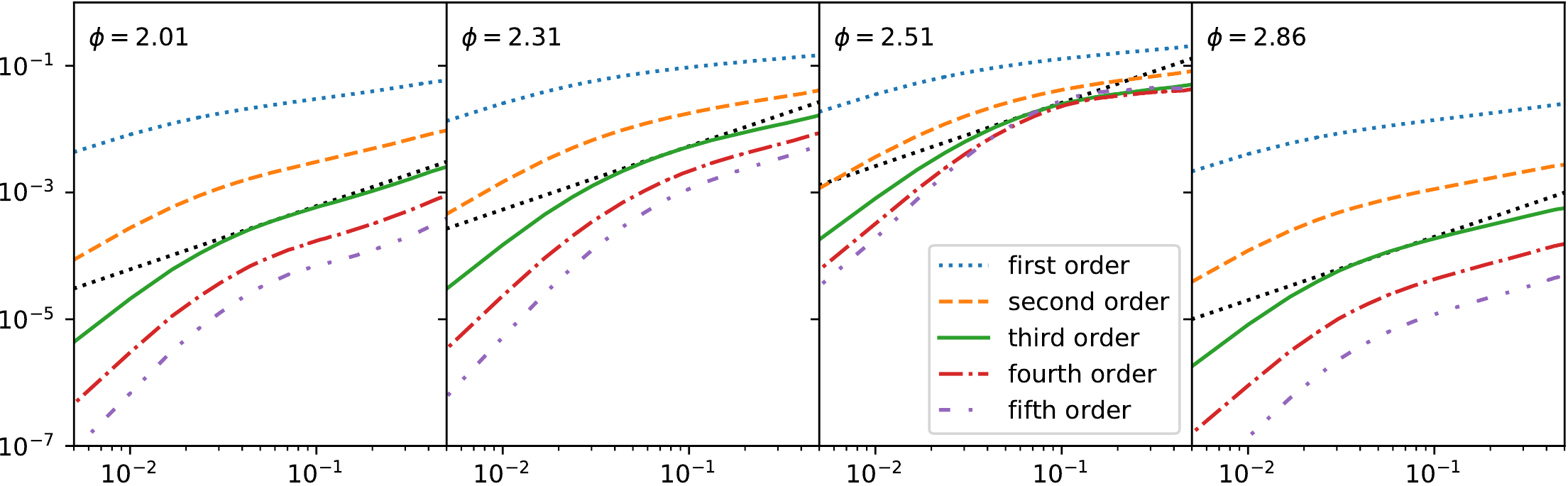}
	\end{picture}
	\caption{\label{FigStructFuncs}
	Structure functions at four different orbital phases as indicated in the figures.
	For comparison, we show a linear function by the black dotted line.
	By our normalisation, the structure functions are given in units of $c^p$.
	}
\end{figure*}

Examples for corresponding structure functions are shown in Fig. \ref{FigStructFuncs}.
We observe only a tiny inertial range, if any, at scales around 0.07 AU, where $S_3$ in most phases shows a linear dependence on $l$.
Especially for phases just around apastron, the structure functions often show rather erratic behaviour, where we give one of the more well-behaved examples in Fig. \ref{FigStructFuncs}.

In contrast to homogeneous turbulence simulations, where structure functions are often analysed and show typical behaviour given by the dissipative structures \citep[][]{SheLeveque1994PhRvL72_336}, the driving in our simulations is not constrained to the largest spatial scales.
Instead it follows from the different instabilities, which drive the fluctuations at a range of spatial scales.
Therefore, we do not have an extended inertial range in our simulations, where only the effect of the turbulent cascade dominates.
Even though most of this driving occurs in the vicinity of the contact discontinuity, the corresponding fluctuations are advected into our analysis regions.
This becomes particularly important for orbital phases around and after apastron, where the region filled by unshocked pulsar wind becomes rather large, thus shifting the region with the turbulent driving closer to the turbulence-analysis region.
In some of these phases additionally, parts of the unshocked pulsar wind extend into the volume, where we analyse the fluctuations. 
Despite all this, the structure functions are related to the amplitude of fluctuations within the orbital domain, where we observe strong differences between the different phases depicted in Fig. \ref{FigStructFuncs}.
This strong variability also observed in the other previous plots, motivates an analysis of the short- and long-term variability of different quantities in this system.

\subsection{Short-Term and Orbit-to-Orbit Variations}
An important property of our new set of simulation results is the long timescale considered, here.
Apart from the initialisation phase in the first orbit, we considered the evolution of the system for two further full orbits. Therefore, we are in a position to quantify dynamical changes during a single orbit as well as orbit-to-orbit variations. 

As a first relevant variable that can be easily obtained from the simulation results, we investigated the size of the volume filled with unshocked pulsar wind. This was obtained by adding the volume of all cells with a four velocity fulfilling $\gamma > 7$.
As can also be seen in Fig. \ref{FigVelocitySnapshots}, outside the unshocked pulsar wind such speeds are not achieved despite the re-acceleration of shocked pulsar wind in the flanks of the WCR.

\begin{figure}
\resizebox{\hsize}{!}{
	\setlength{\unitlength}{0.0081cm}
	\begin{picture}(1050,766)(-50,-50)
		\put(-50,260){\small \rotatebox{90}{$V_{\text{pulsar}}$ [AU$^3$]}}
		\put(500,-50){\small $\Phi$}
		\includegraphics[width=1000\unitlength]{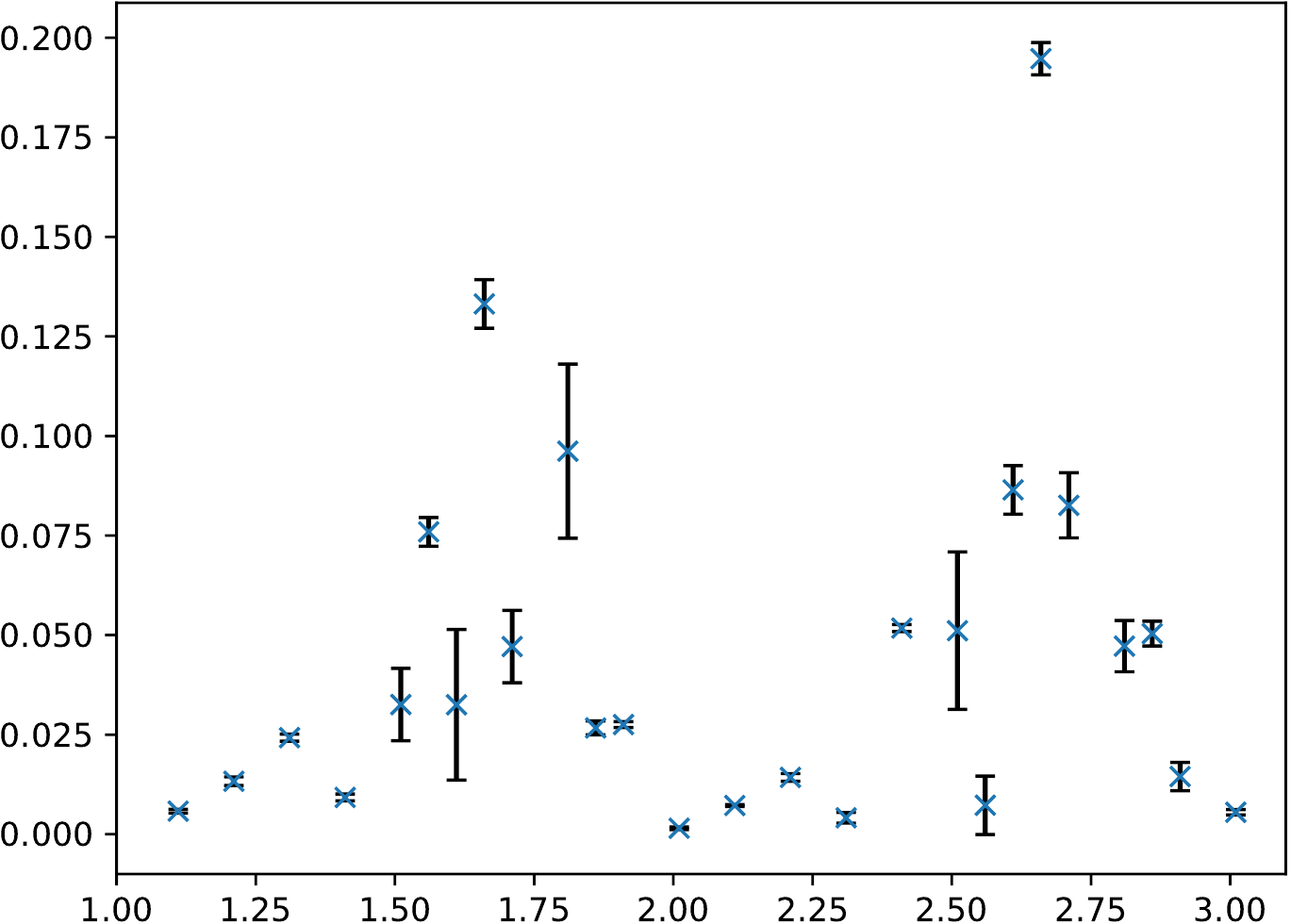}
	\end{picture}
}
\caption{
Volume of unshocked pulsar wind $V_{\text{pulsar}}$ as a function of orbital phase $\Phi$.
Data are shown for the full second and third orbit.
The error bars indicate the corresponding standard deviation, that was computed from multiple output files at similar orbital phases, where we averaged over six phases spread over slightly more than 40 min.}
\label{FigPulsarWindVolume}
\end{figure}

Fig. \ref{FigPulsarWindVolume} shows some remarkable features regarding the stability of the unshocked pulsar-wind structure. First, its size apparently is very stable right after periastron. After that, the size expectedly grows, but the fluctuations become very large, especially after apastron.
As a measure of the fluctuations we computed the standard deviation of volumes extracted from six consecutive output files.
Visual inspection of Fig. \ref{FigPulsarWindVolume} shows that this estimate sometimes underestimates the actual variation in our simulations, where strong changes occur at times, for which we have no output data available.

Considering that this is volume and not linear distance, length scales will show correspondingly smaller variations, but it can still be expected that emission from this system will show significantly higher fluctuations around and in particular after apastron. From the given averaged values we find the smallest volume of the unshocked pulsarwind to be 0.0015 AU$^3$ and the largest, the peak in the third orbit, 0.19 AU$^3$. Converting this to a linear distance, this corresponds to a factor of more than 5 in spatial extent, which correlates nicely to the expectation, when considering Fig. \ref{FigDensOrbit2}.

\begin{figure}
\resizebox{\hsize}{!}{
	\setlength{\unitlength}{0.0081cm}
	\begin{picture}(1050,766)(-50,-50)
		\put(-50,300){\small \rotatebox{90}{$d$ [AU]}}
		\put(500,-50){\small $\Phi$}
		\includegraphics[width=1000\unitlength]{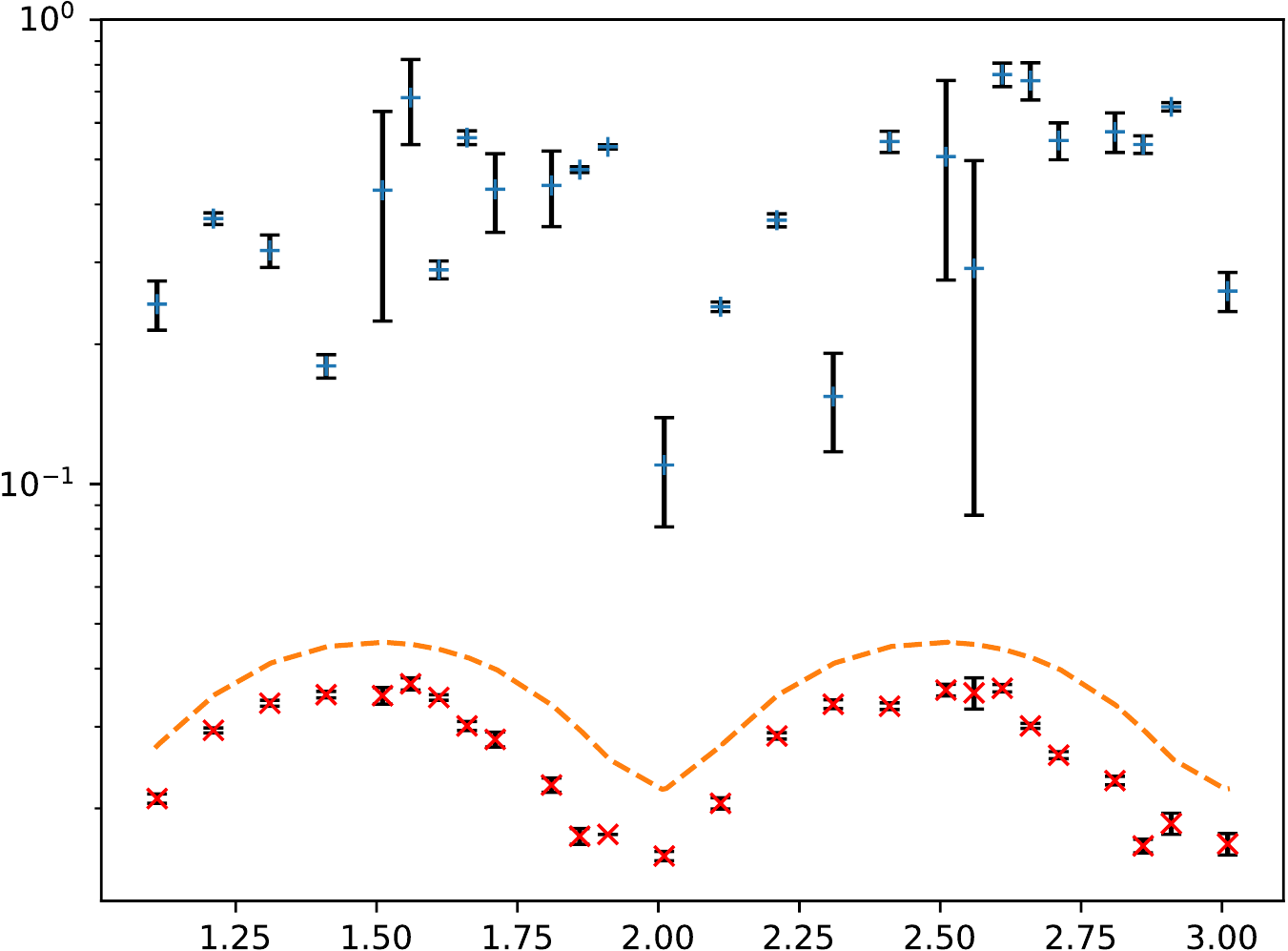}
	\end{picture}
}
\caption{Distance of the Coriolis shock (blue plus signs) and bow shock (red crosses) from the location of the pulsar as functions of orbital phase, where distances are given along the line connecting star and pulsar. Average values and error bars were determined as stated in Fig. \ref{FigPulsarWindVolume}. The $y$-axis uses logarithmic units to allow showing both shock distances simultaneously. Additionally, the dashed orange line shows the approximate distance of the contact discontinuity from the pulsar in the direction of the star using the formula by \citet{EichlerUsov1993ApJ402_271}.}
\label{FigShockDistances}
\end{figure}

In Fig. \ref{FigShockDistances} we show the orbital variation of the distance of bow and Coriolis shocks from the pulsar.
These were computed by following the direction connecting star and pulsar from the pulsar until the Lorentz-factor is smaller than 7.
As expected, the distance of the bow shock from the pulsar varies smoothly over the orbit -- fluctuations of this distance are small and the variation corresponds to the variation of the orbit.
At all phases the distance of the bow shock is at a similar fraction of the approximate distance of the contact discontinuity:
\begin{equation}
	d_{CD} =   \frac{\sqrt{\eta}}{1+\sqrt{\eta}}
\end{equation}
as given by \citet{EichlerUsov1993ApJ402_271}.
In contrast the distance of the Coriolis shock varies significantly, and not always in accordance with the orbit.
Again, strong variations around apastron are present, but also around periastron the variation can be quite large.
This again reflects the sudden displacements of the Coriolis shock as shown in Fig. \ref{FigShortTermVariation} and discussed in Sec. \ref{SecCoriolisShock}.
The implications for gamma-ray emission from the particle population related to the Coriolis shock will be investigated in a future study.

The distribution of the Lorentz factor in Fig. \ref{FigDistGamma} also shows distinct orbital variation -- related to the variation of turbulence.
In this plot, we can also see the variation of the volume of the unshocked pulsar wind, as discussed above, from the changing magnitude of the peak near $\gamma=7$.
The varying fraction of unshocked stellar wind at low gamma factors simply follows through the motion of the stars within the co-rotating domain, which leads to different volumes of unshocked stellar wind as also visible in Fig. \ref{FigDensOrbit2}.
The imprint of turbulence can rather be seen at intermediate gamma factors.
Also here, the changing volume due to the moving stars has some impact, but apart from that especially during the second orbit the  $4 < \gamma < 7$ regime shows changes related to the dynamics of turbulence, mixing, and dynamical changes of the region around the pulsar-wind termination shock.
Apparently, for phases just after apastron, the fraction of shocked pulsar wind in the computational domain is much larger, leading to higher contributions of the intermediate gamma values.
Especially at these phases, the unshocked pulsar-wind volume is rather large with a surrounding region of sometimes smooth flow (see Sec. \ref{SecCoriolisRegion} and the right-handed figure of Fig. \ref{FigVelocitySnapshots}), where reacceleration and shocks with large angles between shock normal and pulsar-wind velocity can lead to high gamma factors in this post-shock region.

This behaviour is qualitatively also visible in the third orbit (see middle plot in Fig. \ref{FigDistGamma}), but it is less pronounced in this case.
This hints at strong orbit-to-orbit variability of the turbulence within this system.
Apart from that, we illustrate the short-time variability of the velocity in the right-handed plot of Fig. \ref{FigDistGamma}.
Apparently, distinct changes of the distribution function of the gamma factor occur even on timescales of a few minutes.
In this case, the distribution decreases with time in the region $4 < \gamma < 7$, in accordance with the observations in Fig. \ref{FigShortTermVariation}, where we saw that during this time a large region filled with a smooth high-gamma-factor wind is disrupted and filled with slower, more turbulent matter.

\begin{figure}
\resizebox{\hsize}{!}{
	\setlength{\unitlength}{0.0081cm}
	\begin{picture}(1050,799)(-50,-50)
		\put(-50,150){\small \rotatebox{90}{fraction of supersonic medium}}
		\put(500,-50){\small $\Phi$}
		\includegraphics[width=1000\unitlength]{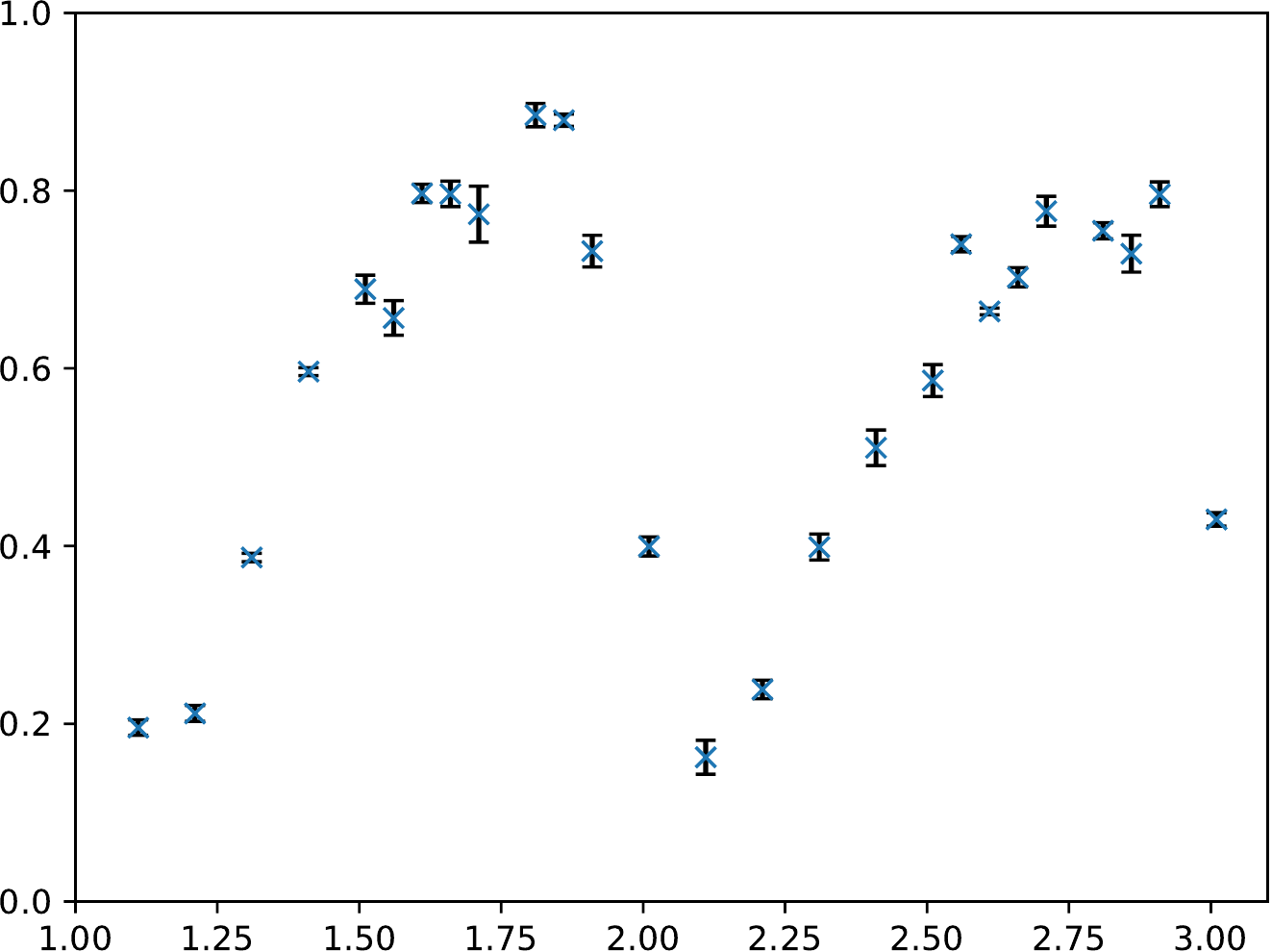}
	\end{picture}
}
\caption{
Fraction of supersonic downstream medium as a function of orbital phase.
Average values and error bars were determined as stated in Fig. \ref{FigPulsarWindVolume}.}
\label{FigFractionSupersonic}
\end{figure}

From the distribution of the Mach number in the computational domain, we additionally computed the fraction of supersonic flow in the downstream region.
For this, the unshocked stellar was identified by its small speed of sound and the unshocked pulsar wind by its large Lorentz factor.
Both unshocked winds were excluded from this analysis since the flow is highly supersonic in each case.
As can be seen in Fig. \ref{FigFractionSupersonic}, the fraction of downstream medium that moves supersonically in our simulation depends strongly on orbital phase, with a clear peak after apastron.
This behaviour is also visible in Fig. \ref{FigMachOrbit2}, where phases near and after apastron feature large regions with high Mach-number flow - especially relating to the smooth flow surrounding the unshocked pulsar wind.

Since neither the mere amplitude of the velocity as given by the distribution of the gamma factor nor the Mach number of the flow do directly give the turbulence amplitude, we additionally computed an estimate for the mean rate of inertial energy dissipation from the third-order longitudinal structure functions $S_3$.
Using Kolmogorov's four-fifth law:
\begin{equation}
S_3 = - \frac{4}{5} \epsilon l,
\end{equation}
$\epsilon$ was computed by integrating $S_3$ over the range of scales, where $S_3$ follows the expected $S_3 \propto l$ dependence (see Fig. \ref{FigStructFuncs}), and then solving for $\epsilon$.
Its orbital dependence is similar to the previously discussed ones, with peaks around periastron and strong variation around the same time.
In particular around apastron, we observed several phases, where the region in which we evaluate the structure functions is permeated by parts of the unshocked pulsar wind.
This leads on the one hand to non-turbulent regions within the analysis volume.
On the other hand the contact discontinuity is close by the analysis region and is part of it in some cases (more precisely the phases near $\Phi=1.4-1.6, 2.55-2.6$ are affected).
This also means that the instability driving the turbulence is active within the analysis volume at scales, which are investigated for fully developed turbulence.
As a result, the structure functions are not representative for fully developed turbulence.
Nonetheless, in nearly all cases, we find $S_3 \propto l$ in the regime from which we compute $\epsilon$.
Therefore, we still use the result as a measure of the fluctuations inside the relevant volume.
Especially for the phases around apastron we observe strong short-term fluctuations in $\epsilon$.

\begin{figure}
\resizebox{\hsize}{!}{
	\setlength{\unitlength}{0.0081cm}
	\begin{picture}(1050,816)(-50,-50)
		\put(-50,380){\small \rotatebox{90}{$\epsilon$}}
		\put(500,-50){\small $\Phi$}
		\includegraphics[width=1000\unitlength]{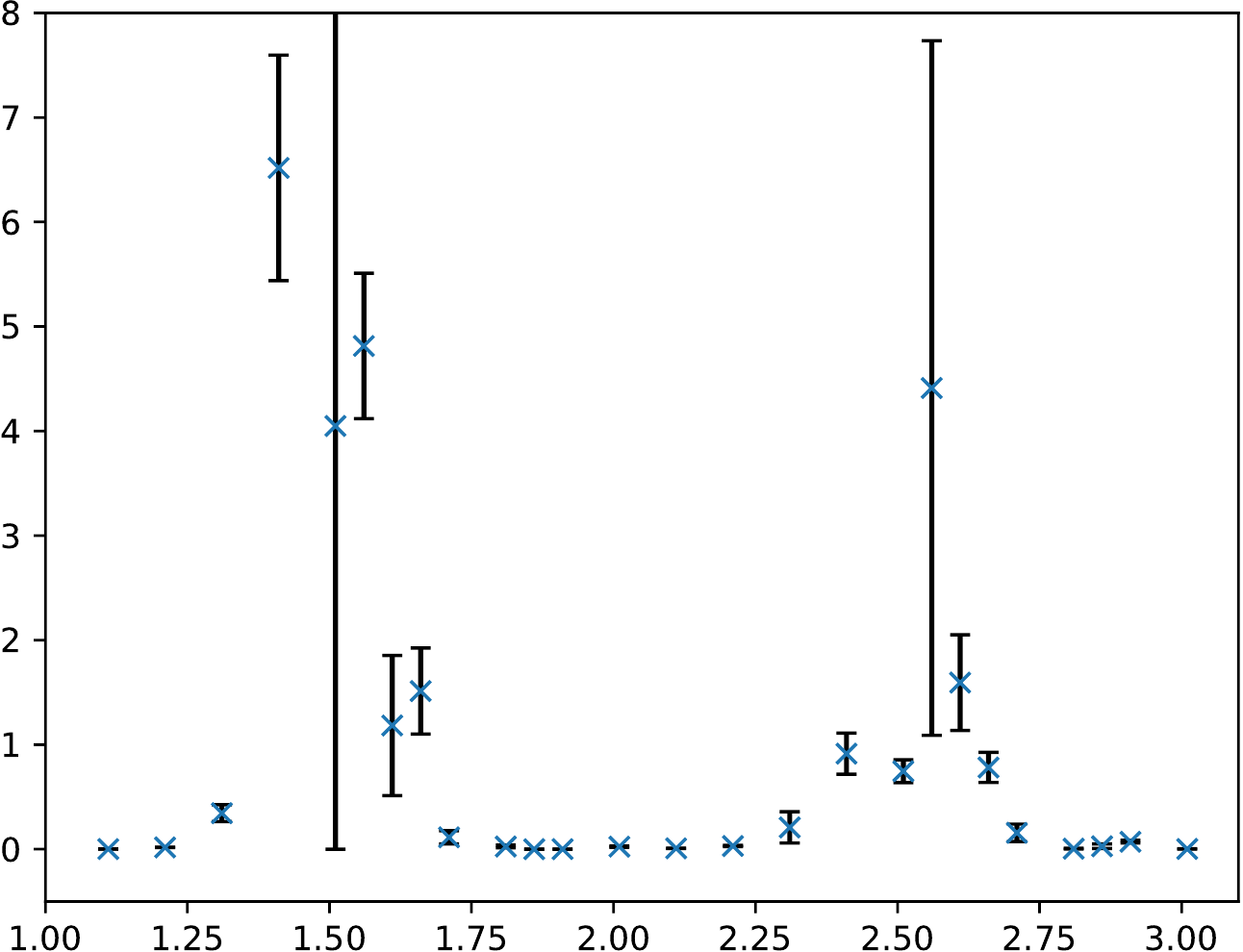}
	\end{picture}
}
\caption{Variation of the mean rate of inertial energy dissipation per unit mass $\epsilon$ as a function of orbital phase $\Phi$ over the second and the third orbit.
As in Fig. \ref{FigPulsarWindVolume}, the error bars indicate the standard deviation computed from six consecutive output files.
$\epsilon$ is given in units of $c^3$/AU.}
\label{FigTurbOrbit}
\end{figure}

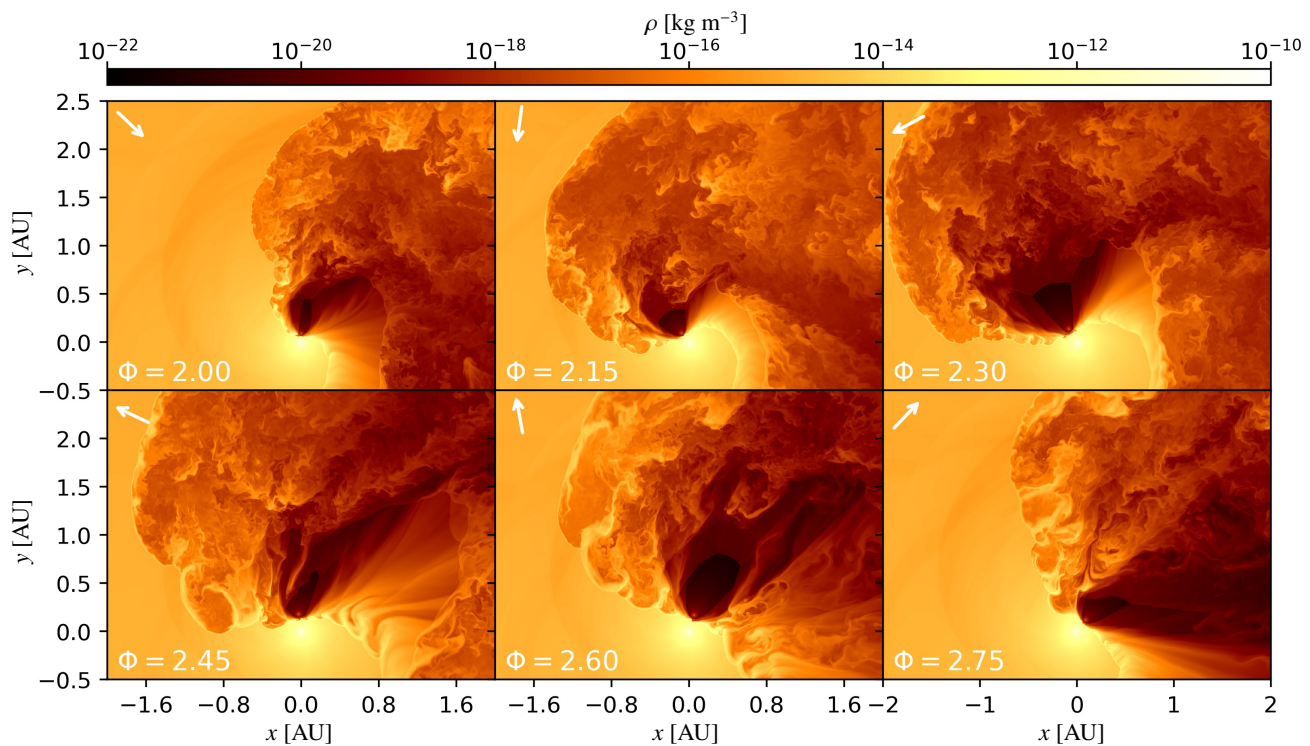
\begin{figure*}
\centering
\setlength{\unitlength}{0.0166666cm}
\begin{picture}(1020,598)(-20,-30)
\put(-20,120){\small \rotatebox{90}{$y$ [AU]}}
\put(-20,350){\small \rotatebox{90}{$y$ [AU]}}
\put(180,-20){\small $x$ [AU]}
\put(485,-20){\small $x$ [AU]}
\put(790,-20){\small $x$ [AU]}
\put(480,545){\small $\rho$ [kg m$^{-3}$]}
\includegraphics[width=1000\unitlength]{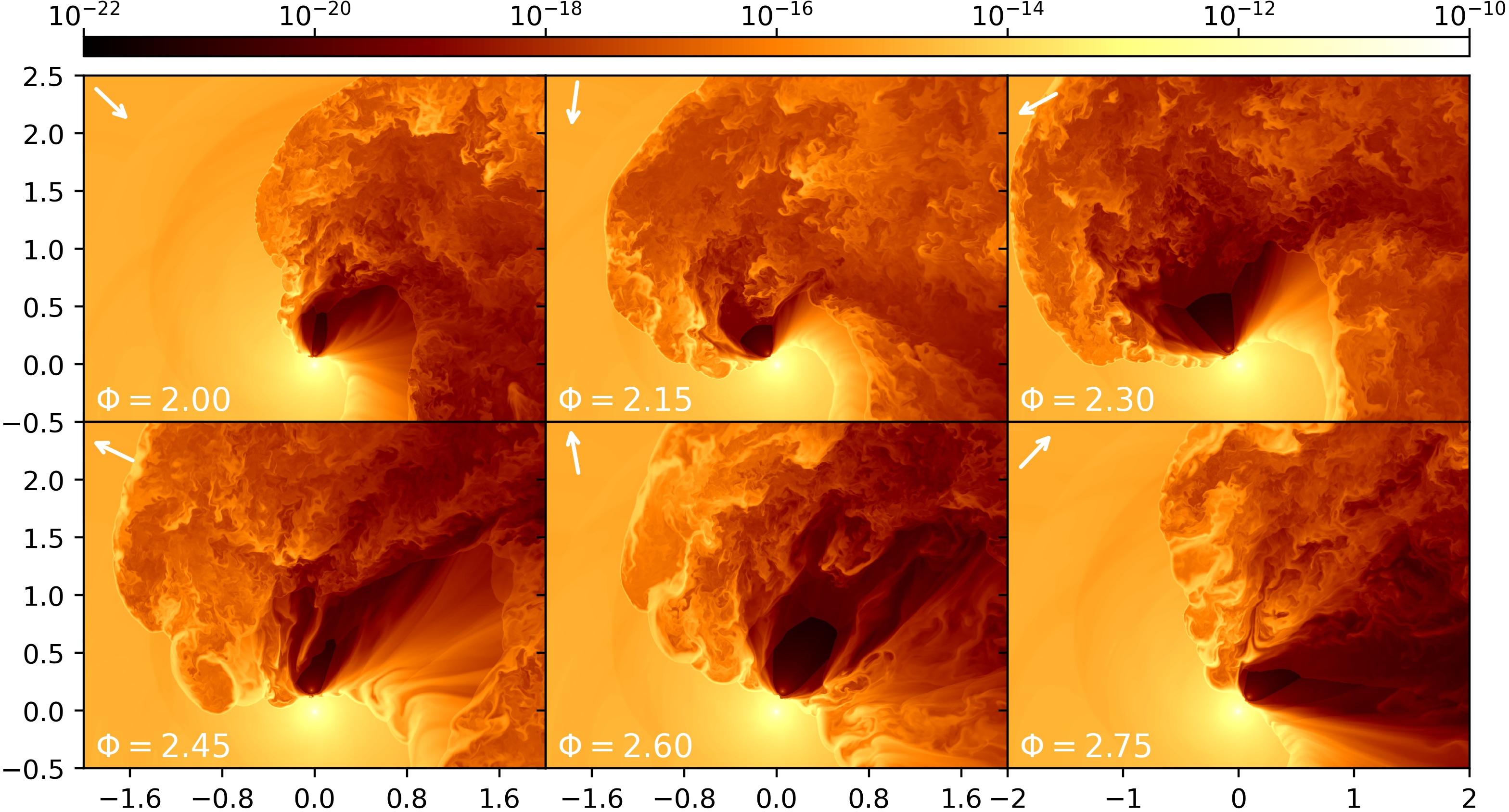}
\end{picture}
\caption{Same as Fig. \ref{FigDensOrbit2} but for the third full orbit.}
\label{FigDensOrbit3}
\end{figure*}

Apart from the short-time variation discussed so far, we also observe orbit-to-orbit variations.
In Figs. \ref{FigPulsarWindVolume} - \ref{FigTurbOrbit}, we find that the phase dependence around periastron is very similar in the second and the third orbit.
Around apastron, however, the short-term fluctuations are stronger than the orbital ones leading to different phase dependencies in the two investigated orbits.
This can also be seen, when comparing the mass density in the orbital plane in the third orbits shown in Fig. \ref{FigDensOrbit3} to the one in the second orbit shown in Fig. \ref{FigDensOrbit2}.
Especially for the shape of the unshocked pulsar-wind region and the immediate surroundings we find strong differences especially after apastron -- see the density distributions at $\Phi=1.6$ vs. $\Phi=2.6$ or at $\Phi=1.75$ vs. $\Phi=2.75$.
This corresponds to the times, when we observe large fluctuations in the volume of the unshocked pulsar wind in Fig. \ref{FigPulsarWindVolume}.

This might also have implications for the variability of non-thermal radiation observed from the system.
From our investigations of the dynamics of the system, we expect stable gamma-ray emission around periastron with possibly strong variations around apastron.
\citet{HuberEtAl2021AnA649_71} found that the gamma-ray emission at the time shortly after apastron is particularly affected by relativistic beaming related to the emission of energetic particles within the leading edge of the shocked pulsar wind.
From Figs. \ref{FigDensOrbit2} and \ref{FigDensOrbit3}, we see that the flow direction in this region is aligned with the direction to the observer some time around relative phase $\Phi=0.6$, where in the current simulation this situation seems to occur a little later than in the simulations by \citet{HuberEtAl2021AnA649_71}.
However, as also discussed in \citet{HuberEtAl2021AnA649_71} this phase was not well represented in their simulations due to the limited size of their numerical domain.

This relativistic beaming occurs just in the region, where we have the rather laminar flow within the leading edge of the shocked pulsar wind, which is also visible as the dark yellow region in Fig. \ref{FigVelocitySnapshots}.
That same figure shows that at some phases, there even appear two distinct such regions with different overall flow directions.
Regarding the impact on gamma-ray emission, only the relative phases around $\Phi=0.6$ when the flow is aligned with the observer direction will be important.
Due to the rather different structure of this region at phases $\Phi=1.6$ and $\Phi=2.6$, however, we can expect to find a different impact of relativistic beaming for the gamma-ray emission in the second and the third orbit.
This will be studied in more detail in a forthcoming publication, where we will investigate the propagation of energetic particles within the simulated stellar winds, discussed here.
There, it will be particularly interesting if we find short-term and orbit-to-orbit variations on a similar level as we observe in the present study.

\section{Summary and Discussion}
\label{SecSummary}
In this study, we investigated the dynamical interaction of the pulsar wind and the stellar wind in the gamma-ray binary system LS~5039 via RHD simulations, where we assumed the wind-driven scenario to explain the observed gamma-ray emission.
Here, we did not take the magnetic field in the winds into account, where previous 2D simulations led to the expectation of a rather low impact on the wind dynamics \citep[see the discussion in][]{BogovalovEtAl2012MNRAS419_3426,Bosch-RamonEtAl2015AnA577_89}. 
Recent 3D RMHD simulations \citep[][]{BarkovEtAl2022arXiv221112053}, however, show that for future simulations it will also be interesting to include the effects of the magnetic field. 
The investigated simulation was done with unprecedented resolution over three full orbits, to allow a detailed analysis of the turbulence driven by the wind-wind interaction in the WCR together with the short-term and long-term variability of the wind dynamics.

In the simulation, we observe strong turbulence in the downstream region of the WCR, where we do not see a clear, broad inertial range due to our driving force by the different instabilities being active at a range of spatial scales.
Here, we did not take clumping for the stellar wind into account, which can be shown to further increase fluctuations \citep[][]{Paredes-Fortuny2015AnA574_77, KefalaBosch-Ramon2023AnA669_21}.
Thus, short-term and orbit-to-orbit variation might actually be even stronger than found in our model.
However, one has to be aware that the Lorentz factor of the pulsar wind, chosen in our and similar simulations, was significantly smaller than expected in a realistic pulsar wind.
While this is expected to diminish growth rates for the KH instability\citep[][]{FerrariEtAl1980MNRAS193_469}, the RM and RT instabilities, which seem to be responsible for a large part of the observed turbulence should be much less affected by a larger Lorentz factor and a corresponding change in density contrast \citep{AllenHughes1984MNRAS208_609, Bosch-RamonEtAl2015AnA577_89}.

In our configuration we find strong variability due to the turbulence driven within the WCR.
While different parameters turn out to be stable around periastron, we observe particularly strong fluctuations both on short and on orbit-to-orbit timescales around apastron. 
These fluctuations seem to be stronger than found in previous simulations as, e.g., in \citet{Bosch-RamonEtAl2015AnA577_89}.
This might be partly attributed to the higher resolution especially in the outer parts of the numerical domain, but also to the somewhat larger eccentricity used in our simulation.

In a future analysis, we will further investigate how the strong dynamics observed here will impact energetic particle transport and ensuing emission of non-thermal radiation from this system.
Again the phase shortly after apastron will be particularly interesting, because at this time we observed the largest fluctuations, while it is also the time, where relativistic beaming in the direction of the observer can enhance emission from within the WCR.

\begin{acknowledgements}
	We thankfully acknowledge PRACE for granting us access to Joliot-Curie at GENCI@CEA, France. 
	We thankfully acknowledge the access to the research infrastructure of the Institute for Astro- and Particle Physics at the University of Innsbruck (Server Quanton AS-220tt-trt8n16-g11 x8).
  This research made use of Cronos \citep{KissmannEtAl2018ApJS236_53}; GNU Scientific Library (GSL) \citep{galassi2018scientific}; FFTW3 \citep[][]{FrigoJohnson2005ProcIEEE93_216}, matplotlib, a Python library for publication quality graphics \citep{Hunter2007}; Scipy \citep{2020SciPy-NMeth}; and NumPy \citep{2020NumPy-Array}.
\end{acknowledgements}

\bibliographystyle{aa}
\bibliography{referencesRHD}
\newpage
\begin{appendix}
\section{Supplementary material}
For further illustration of the dynamics of the system in our simulation, we supply a video of the evolution of the mass density in the orbital plane of the LS-5039 system.
The video is produced from the series of six successive outputs for different orbital phases during the full second and third orbit.
Since these outputs are not homogeneously distributed over all phases of the orbit, the video continuously jumps from phase to phase, where it shows a brief moment of the short-time dynamics of the system.

This video is useful in showing several of the effects discussed in this paper:
the vast difference in speed of the stellar and the pulsar wind becomes directly obvious from the very different short-time motion of the material.
Additionally, the highly dynamical changes of the shape and the size of the volume filled by unshocked pulsar wind, becomes very apparent.
Finally, only such a video can clearly show the large-scale motion of the material over longer timescales.
\end{appendix}

\end{document}